\documentclass[a4paper,11pt]{article}
\usepackage{pos}

\newcommand{\be}{\begin{equation}}
\newcommand{\ee}{\end{equation}}
\newcommand{\bea}{\begin{eqnarray}}
\newcommand{\eea}{\end{eqnarray}}

\title{Exclusive B-meson semileptonic decays from unitarity and lattice QCD}

\author[a]{G.~Martinelli}
\author[b]{M.~Naviglio}
\author*[c]{S.~Simula}
\author[d]{L.~Vittorio}

\affiliation[a]{Physics Department, University of Roma ``La Sapienza'' and INFN, Sezione di Roma,\\ Piazzale Aldo Moro 5, 00185 Roma, Italy}

\affiliation[b]{Dipartimento di Fisica dell'Universit\`a di Pisa and INFN, Sezione di Pisa,\\ Largo B.~Pontecorvo 3, I-56127 Pisa, Italy}

\affiliation[c]{Istituto Nazionale di Fisica Nucleare, Sezione di Roma Tre,\\ Via della Vasca Navale 84, I-00146 Rome, Italy}

\affiliation[d]{Scuola Normale Superiore, Piazza dei Cavalieri 7, 56126 Pisa, Italy and\\INFN, Sezione di Pisa, Largo B.~Pontecorvo 3, I-56127 Pisa, Italy}

\emailAdd{guido.martinelli@roma1.infn.it}
\emailAdd{manuel.naviglio@phd.unipi.it}
\emailAdd{simula@roma3.infn.it}
\emailAdd{ludovico.vittorio@sns.it}

\abstract{We examine the semileptonic $B \to D^{(*)} \ell \nu_\ell$ and $B \to \pi \ell \nu_\ell$ decays adopting the unitarity-based Dispersive Matrix (DM) method, which allows to determine the shape of the relevant hadronic form factors (FFs) in their whole kinematical range, using only lattice QCD results available at large values of the 4-momentum transfer without making any assumption on their momentum dependence. Moreover, the experimental data are not used to constrain the shape of the FFs, but only to obtain our final exclusive determination of $\vert V_{cb} \vert$  and $\vert V_{ub} \vert$, namely: $\vert V_{cb} \vert \cdot 10^3 = 41.1 \pm 1.0$ and $\vert V_{ub} \vert \cdot 10^3 = 3.88 \pm 0.32$, which are consistent with the latest inclusive determinations at the $1 \sigma$ level or better. 
Our calculation of the FFs allows to obtain pure theoretical estimates of the $\tau / \mu$ ratios of differential decay rates, $R(D) = 0.296 \pm 0.008$ and $R(D^*) = 0.275 \pm 0.008$, which turn out to be compatible with the experimental world averages within $\simeq 1.4$ standard deviations.}

\FullConference{%
  11th International Workshop on the CKM Unitarity Triangle (CKM2021)\\
  22-26 November 2021\\
  The University of Melbourne, Australia
}

\begin{document}
\maketitle

\section{Introduction}

The exclusive semileptonic $B \to D^{(*)} \ell \nu_\ell$ decays are very intriguing processes from a phenomenological point of view, mainly for two reasons. 
The first one is the $\vert V_{cb} \vert$ puzzle, $i.e.$ the tension between the inclusive and exclusive determinations of the Cabibbo-Kobayashi-Maskawa (CKM) matrix element $|V_{cb}|$, which, according to the latest version of the FLAG review\,\cite{Aoki:2021kgd}, is at the level of $\simeq 2.7$ standard deviations. 
The second reason is the discrepancy between the Standard Model (SM) predictions and the experiments in the determination of the $\tau / \mu$ ratios of the branching fractions, $R(D^{(*)})$, which represent a fundamental test of Lepton Flavour Universality (LFU) in the SM. According to the HFLAV Collaboration\,\cite{HFLAV:2019otj} the above discrepancy is at the level of $\simeq 3.1 \sigma$.

In addition a long-standing tension has affected also the inclusive and the exclusive determinations of the CKM matrix element $\vert V_{ub} \vert$ until a recent measurement of the inclusive value of $\vert V_{ub} \vert$ made by Belle~\cite{Belle:2021eni} has changed the picture. The last PDG review\,\cite{ParticleDataGroup:2020ssz} quotes a discrepancy of $\simeq 1.4\sigma$.

In this contribution our aim is to examine the semileptonic $B \to D^{(*)} \ell \nu_\ell$  and $B \to \pi \ell \nu_\ell$ decays adopting the unitarity-based Dispersive Matrix (DM) method of Ref.\,\cite{DiCarlo:2021dzg}, which can be applied to any semileptonic decays once lattice QCD (LQCD) computations of the relevant susceptibilities and of the form factors (FFs) are available. 
Only LQCD computations of the FFs at large values of the 4-momentum transfer will be used to determine the shape of the FFs in the whole kinematical range without making any assumption on their momentum dependence. Moreover, the experimental data are not used to constrain the shape of the FFs, but only to obtain the final exclusive determination of $\vert V_{cb} \vert$ and $\vert V_{ub} \vert$. In this way, our determination of the FFs allows to obtain pure theoretical estimates of several quantities of phenomenological interest, namely the $\tau / \mu$ ratios of differential decay rates, various polarization observables and forward-backward asymmetries.

\section{The DM method}

We now briefly recall the main features of the DM method applied to the description of a generic form factor $f(q^2)$ relevant in the decay between hadrons with mass $m_1$ and $m_2$\,\cite{DiCarlo:2021dzg}.
Given a set of known values of the form factor, i.e.~$\{ f_j \equiv f(q_j^2) \}$ with $j = 1, 2, ..., N$, and of the corresponding susceptibility $\chi$, the form factor at a generic value of $q^2$ is bounded by unitarity, analyticity and crossing symmetry to be in the range
\be
  \beta(z) - \sqrt{\gamma(z)} \leq f(z) \leq \beta(z) + \sqrt{\gamma(z)} ~ , ~
    \label{eq:bounds}
\ee 
where 
\bea
      \label{eq:beta_final}
      \beta(z) & \equiv & \frac{1}{\phi(z) d(z)} \sum_{j = 1}^N \phi_j f_j d_j \frac{1 - z_j^2}{z - z_j} ~ , ~ \\
      \label{eq:gamma_final}
      \gamma(z) & \equiv &  \frac{1}{1 - z^2} \frac{1}{\phi^2(z) d^2(z)} \left( \chi - \chi_{DM} \right) ~ , ~ \\
      \label{eq:chiDM}
      \chi_{DM} & \equiv & \sum_{i, j = 1}^N \phi_i f_i \phi_j  f_j d_i d_j \frac{(1 - z_i^2) (1 - z_j^2)}{1 - z_i z_j} ~ 
\eea
with $ d(z) \equiv \prod_{m = 1}^N (1 - z z_m) / (z - z_m)$ and $ d_j  \equiv \prod_{m \neq j = 1}^N (1 - z_j z_m) / (z_j - z_m)$.
In Eqs.\,(\ref{eq:bounds})-(\ref{eq:chiDM}) $z = (\sqrt{t_+ - q^2} - \sqrt{t_+ - t_-}) / (\sqrt{t_+ - q^2} + \sqrt{t_+ - t_-}))$ is the conformal variable, $t_\pm \equiv (m_1 \pm m_2)^2$ and the quantities $\phi_j \equiv \phi(z_j)$ are the values of the kinematical function appropriate for the given form factor\,\cite{Boyd:1997kz} containing the contribution of the resonances below the pair production threshold $t_+$.

Unitarity is satisfied only when $\gamma(z) \geq 0$, which implies $\chi \geq \chi_{DM}$.
Since $\chi_{DM}$ does not depend on $z$, the above condition is either never verified or always verified for any value of $z$.
This means that the unitarity filter $\chi \geq \chi_{DM}$ represents a parameterization-independent test of unitarity for a given set of input values $f_j$ of the FF.

We remind an important feature of the DM approach.
When $z$ coincides with one of the data points, i.e.~$z \to z_j$, one has $\beta(z) \to f_j$ and $\gamma(z) \to 0$.
In other words the DM method reproduces exactly the given set of data points.
This is at variance with what may happen using truncated parameterisations based on the $z$-expansion\,\cite{Boyd:1997kz}, since there is no guarantee that such truncated parameterizations reproduce exactly the set of input data.
Thus, it is worthwhile to highlight the following important feature: the DM band given in Eq.~(\ref{eq:bounds}) is equivalent to the results of all possible fits which satisfy unitarity and at the same time reproduce exactly the input data.

\section{Semileptonic $B \to D^{(*)} \ell \nu_\ell$ decays}

The DM method has been applied to the study of the $B \to D \ell \nu_\ell$ decays in Ref.\,\cite{Martinelli:2021onb} and of the $B \to D^* \ell \nu_\ell$ decays in Ref.\,\cite{Martinelli:2021myh}. 
The hadronic FFs are taken from the lattice results of the FNAL/MILC Collaboration\,\cite{MILC:2015uhg,FermilabLattice:2021cdg}, while for the susceptibilities we make use of their nonperturbative computations made on the lattice in Ref.\,\cite{Martinelli:2021frl}.

Using the experimental data from the Belle Collaboration\,\cite{Belle:2015pkj,Belle:2017rcc,Belle:2018ezy} we perform a bin-per-bin determination of $\vert V_{cb} \vert$ by dividing the experimental measurements with the theoretical predictions based on the the parameterization-independent shape obtained with our DM method.
Note that in the case of the $B \to D^* \ell \nu_\ell$ decays we develop an important, specific correction of the experimental correlation matrix of the data of Ref.\,\cite{Belle:2017rcc} (see Refs.\,\cite{Martinelli:2021onb,Martinelli:2021myh}).
We get $\vert V_{cb} \vert \cdot 10^3 = (41.0 \pm 1.2)$ from $B \to D \ell \nu_\ell$ decays and $\vert V_{cb} \vert \cdot 10^3 = (41.3 \pm 1.7)$ from $B \to D^* \ell \nu_\ell$ decays. A simple average of the two results reads
\be
     \label{eq:Vcb}
     \vert V_{cb} \vert \cdot 10^3 = (41.1 \pm 1.0) ~ , ~
\ee
which is compatible with the most recent inclusive result $\vert V_{cb} \vert_{\rm{incl}} \cdot 10^3 = 42.16 \pm 0.50$\,\cite{Bordone:2021oof} at the $\simeq 0.6 \sigma$ level. 
Our determination of the FFs in the whole kinematical range allows us to evaluate a pure, theoretical estimate of the $\tau / \mu$ ratios of the branching ratios, namely
\be
    \label{eq:ratios}
    R(D) = 0.296 \pm 0.008 ~ , ~ \qquad R(D^*) = 0.275 \pm 0.008 ~ , ~
\ee 
which are compatible with the experimental world averages $R(D) = 0.340 \pm 0.027 \pm 0.013$ and $R(D^*) = 0.295 \pm 0.011 \pm 0.008$ from HFLAV\,\cite{HFLAV:2019otj} at the $\simeq 1.4\sigma$ level.

\section{Semileptonic $B \to \pi \ell \nu_\ell$ decays}

In Ref.\,\cite{Martinelli:2022tte} the DM method has been applied to the study of the $B \to \pi \ell \nu_\ell$ decays by evaluating the relevant susceptibilities from suitable two-point correlation functions ad adopting for the hadronic FFs the lattice results of RBC/UKQCD\,\cite{Flynn:2015mha} and FNAL/MILC\,\cite{FermilabLattice:2015mwy} Collaborations at large values of the 4-momentum transfer.

For the extraction of $\vert V_{ub} \vert$ we have analyzed the measurements of six experiments\,\cite{BaBar:2010efp,Belle:2010hep,BaBar:2012thb,Belle:2013hlo}, adopting a bin-per-bin determination in which the experimental data are not used to constrain the shape of the hadronic FFs in order to avoid possible biases.
We get the result
\be
     \label{eq:Vub_BPi}
     \vert V_{ub} \vert \cdot 10^3 = (3.62 \pm 0.47) \qquad \mbox{from~} B \to \pi \ell \nu_\ell ~ \mbox{decays\,} \text{\cite{Martinelli:2022tte}} ~ 
\ee
consistent with the latest inclusive determination $\vert V_{ub} \vert_{incl} \cdot 10^3 = 4.13 \, (26)$ from PDG\,\cite{ParticleDataGroup:2020ssz} at the $\simeq 1\sigma$ level.
It is also compatible with the latest exclusive determination $\vert V_{ub} \vert_{excl} \cdot 10^3 = 3.70\, (16)$\,\cite{ParticleDataGroup:2020ssz} within our larger uncertainty related both to the long extrapolation from the high-$q^2$ region of the input lattice data down to $q^2 = 0$ and to the fact that we do not use the experimental data to constrain the shape of the FFs. 

\begin{figure}[htb!]
\centering{\includegraphics[scale=0.65]{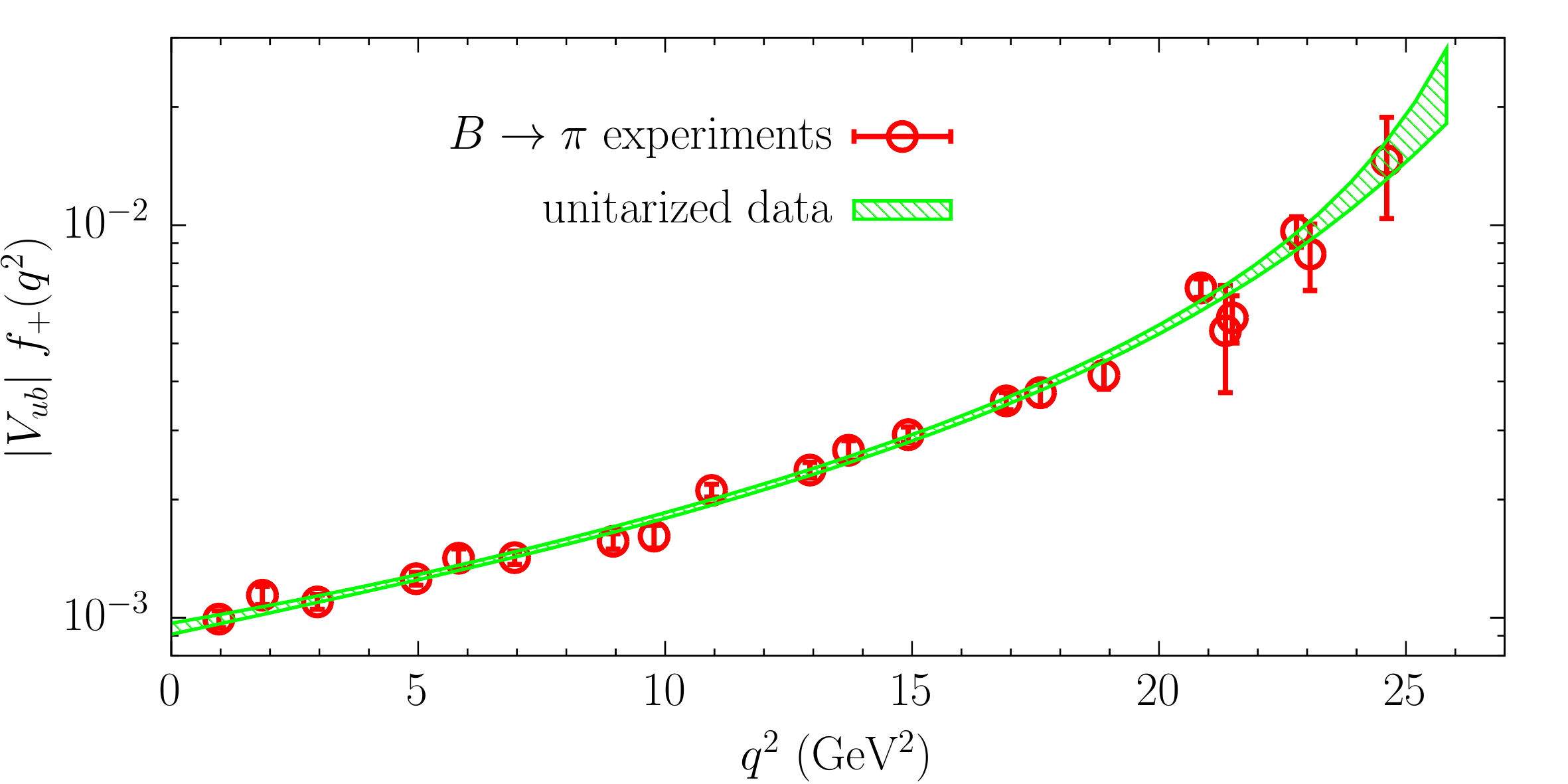}}
\caption{\it \small The values of $\vert V_{ub}  \vert f_+^{B\pi}(q^2)$ for various $q^2$-bins determined from the experimental measurements of Refs.\,\cite{BaBar:2010efp,Belle:2010hep,BaBar:2012thb,Belle:2013hlo} and the results of the unitarization procedure (green band) described in the text.}
\label{fig:bounds_all}
\end{figure}

In order to improve the precision we have tried a novel strategy, namely the {\it unitarization} of the experimental data, which we now synthetically describe. In the limit of massless leptons the measurements of the differential decay rate can be easily transformed into experimental determinations of the product $\vert V_{ub}  \vert f_+^{B\pi}(q^2)$ in various distinct $q^2$-bins, where $f_+^{B\pi}(q^2)$ is the semileptonic vector FF of the $B \to \pi$ transition. To such data we can apply the unitarity filter of the DM method using an initial guess for $\vert V_{ub} \vert$. The filter select only the combinations of the data for the various $q^2$-bins which satisfy unitarity. Then, a new value of $\vert V_{ub} \vert$ is extracted using the hadronic FF $f_+(q^2)$ of the DM method and the procedure is iterated until convergence for $\vert V_{ub} \vert$ is reached.
The results of the {\em unitarization} of the experimental data are shown in Fig.\,\ref{fig:bounds_all}.
The extracted value of $\vert V_{ub} \vert$ is
\be
     \label{eq:Vub_BPi_unitarized}
     \vert V_{ub} \vert \cdot 10^3 = (3.88 \pm 0.32) ~ , ~ 
\ee
which is consistent with the previous finding\,(\ref{eq:Vub_BPi}), but it improves the uncertainty by $\simeq 30 \%$. 

\begin{figure}[htb!]
\centering{\includegraphics[scale=0.80]{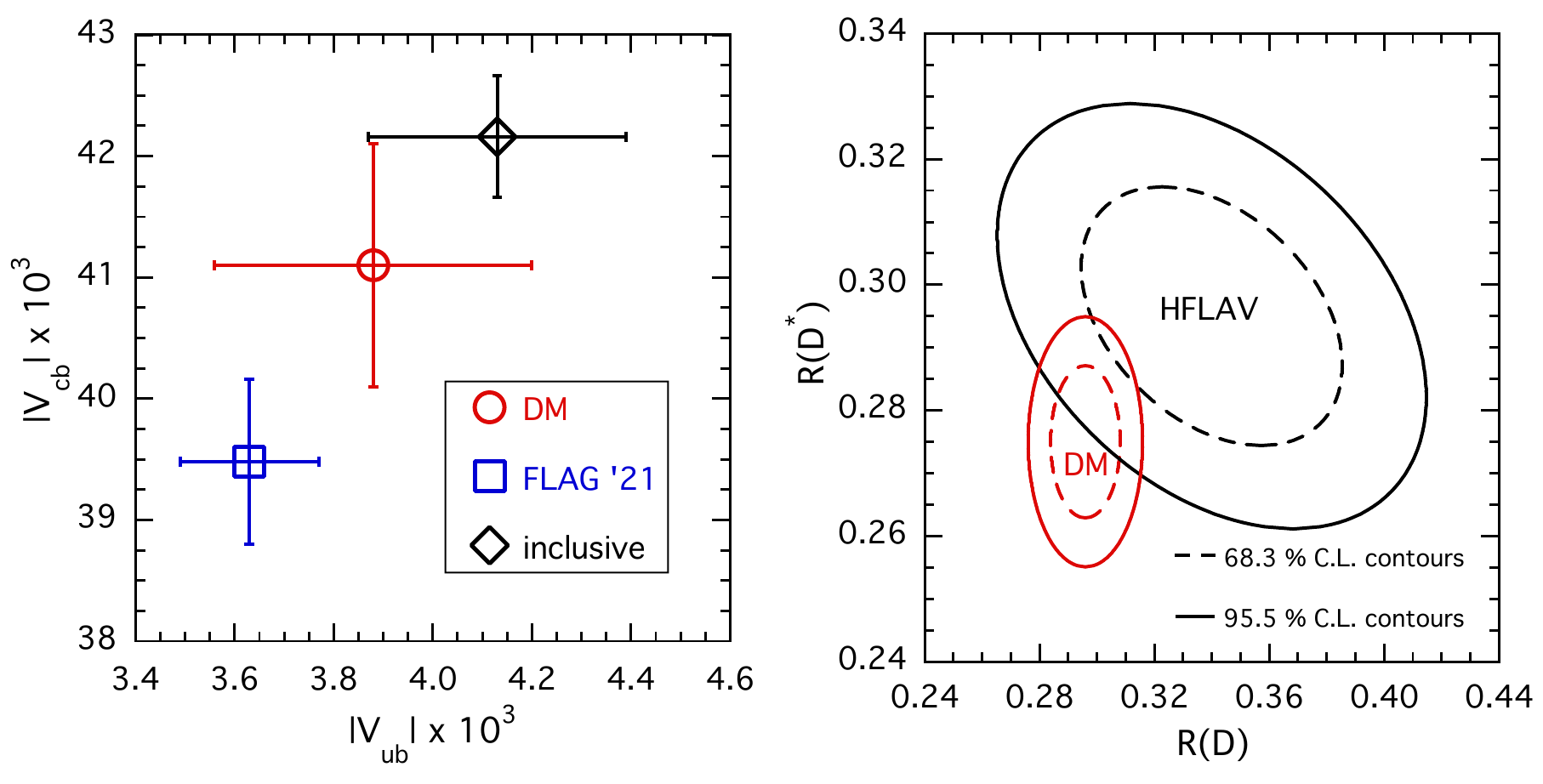}}
\caption{\it \small Left panel: values of $\vert V_{cb}  \vert$ and $\vert V_{ub} \vert$ obtained within the DM method from the analysis of the exclusive $B \to D^{(*)} \ell \nu_\ell$ and $B \to \pi \ell \nu_\ell$ decays compared with the corresponding results quoted in the last FLAG report\,\cite{Aoki:2021kgd} and with the latest inclusive determinations from Refs.\,\cite{ParticleDataGroup:2020ssz,Bordone:2021oof}. Right panel: theoretical estimates of the ratios $R(D^{(*)})$ obtained with the DM method compared with the latest experimental averages from HFLAV\,\cite{HFLAV:2019otj}.}
\label{fig:VcbVub}
\end{figure}

In conclusion, we have analyzed the semileptonic $B \to D^{(*)} \ell \nu_\ell$ and $B \to \pi \ell \nu_\ell$ decays adopting the unitarity-based DM method, which allows to determine the shape of the relevant hadronic FFs in their whole kinematical range using only LQCD results available at large values of the 4-momentum transfer without making any assumption on their momentum dependence. Moreover, the experimental data are not used to constrain the shape of the FFs, but only to obtain our final exclusive determinations of $\vert V_{cb} \vert$ and $\vert V_{ub} \vert$. This allows to obtain pure theoretical estimates of the $\tau / \mu$ ratios of differential decay rates, $R(D)$ and $R(D^*)$. Our findings are collected in Fig.\,\ref{fig:VcbVub} and clearly indicate a significative reduction of the tensions between the exclusive SM predictions with the corresponding inclusive and experimental averages.

\end{document}